\documentclass{article}

\usepackage[main, final]{neurips_2025}

\usepackage[utf8]{inputenc} 
\usepackage[T1]{fontenc}    
\usepackage{hyperref}       
\usepackage{url}            
\usepackage{booktabs}       
\usepackage{amsfonts}       
\usepackage{nicefrac}       
\usepackage{microtype}      
\usepackage{xcolor}         
\usepackage{tabularx}   

\title{Islamic Chatbots in the Age of Large Language Models}

\author{%
  Muhammad Aurangzeb Ahmad\\
  Department of Computer Science \& Software Engineering\\
  University of Washington Bothell\\
  \texttt{maahmad@uw.edu} \\
}

\begin{document}

\maketitle

\begin{abstract}
Large Language Models (LLMs) are rapidly transforming how communities access, interpret, and circulate knowledge, and religious communities are no exception. Chatbots powered by LLMs are beginning to reshape authority, pedagogy, and everyday religious practice in Muslim communities. We analyze the landscape of LLM powered Islamic chatbots and how they are transforming Islamic religious practices e.g., democratizing access to religious knowledge but also running the risk of erosion of authority. We discuss what kind of challenges do these systems raise for Muslim communities and explore recommendations for the responsible design of these systems.
\end{abstract}

\section{Introduction}
Large Language Models (LLMs) are rapidly transforming how communities access, interpret, and generate knowledge. The integration of LLMs into everyday religious practices is leading to blurring the boundary between human authority and algorithmic production. The intersection of AI and religion has drawn increasing scholarly attention in recent years. Researchers in this area note that technologies are never simply neutral carriers of information but are embedded in a cultural and social context \citep{campbell2021}. Digital platforms impact ritual practice, pedagogy, and authority-making in Islam just like other religions.  There is a long history of new technologies affecting Islamic practices. The Ottoman authorities resisted the use of printing press for two centuries until the eighteenth century, in part due to fears of error in reproducing the Quran and a desire to safeguard the prestige of manuscript culture \citep{messick1993}. In the 1970s and 1980s Islamic religious knowledge started to be curated via cassette tape sermons \citep{hirschkind2006}.  In the 1990s, Quran databases on CD-ROM, online fatwa portals, and early Islamic websites such as IslamOnline and Islam Q\&A radically expanded the availability of religious texts and scholarly opinions \citep{bunt2009}. For many Muslims, this was their first encounter with "Sheikh Google," which blurred the boundaries between authoritative scholarship and lay interpretation.

In the 2010s, Islamic knowledge in the public sphere became further entangled with social media e.g., YouTube khutbahs, and Instagram dawah (preaching), and TikTok fatwas. Finally, the advent of LLMs and Islamic chatbots in the 2020s represents the latest chapter in the use of new technologies to access and interpret Islamic knowledge. Tools such as HUMAIN Chat, Ask AiDeen, SheikhGPT, WisQu, Salam.chat, and Ansari Chat offer believers new ways of interacting with religious texts and traditions. The emergence of Islamic chatbots also represents a qualitative shift from earlier media. Unlike static repositories or broadcast platforms, LLM-based systems generate personalized, conversational responses that can resemble scholarly judgment. This is giving rise to the phenomenon of \emph{algorithmic religious authority} which can be defined as the perceived legitimacy granted to computational systems that generate normative or interpretive religious guidance. In traditional Islamic scholarship, authority historically rests on chains of transmission (\emph{isnad}), juristic methodology (\emph{usul al-fiqh}), and scholarly consensus (\emph{ijma}). LLMs are in the process of disrupting these structures by producing fluent outputs without explicit accountability to these mechanisms. 

The current ecosystem of Islamic chatbots is highly heterogeneous. We propose a taxonomy organized along five axes:

\begin{itemize}
\item \textbf{Governance}: state-backed, commercial, or grassroots/open-source systems.
\item \textbf{Scope}: Quran-only, Hadith-focused, jurisprudential, or general guidance.
\item \textbf{Authority Posture}: Educational disclaimers versus fatwa-like responses.
\item \textbf{Technical Grounding}: Retrieval-augmented generation, fine-tuning, or opaque APIs.
\item \textbf{Sectarian Encoding}: Sunni, Shia, Salafi, Sufi etc..
\end{itemize}

These systems promise to bridge linguistic divides, make Islamic knowledge accessible to non-specialists, and provide always-available companions for spiritual inquiry. However, they also raise important questions: Will reliance on chatbots erode the role of the \textit{ulama}? Will state-backed models consolidate interpretive authority, or will open-source alternatives diversify it? Moreover, biases documented in general-purpose LLMs such as Islamophobic framings in GPT-3 outputs \citep{abid2021} underscore the risks of uncritically deploying such systems in religious contexts.

\section{Related Work}
Early efforts around the use of machine learning in this domain focused on Quranic retrieval and Hadith classification using rule-based or shallow machine learning models \citep{shatnawi2014, hussein2018}. More recent work has retrieval-augmented generation (RAG) pipelines to build domain-specific assistants such as MufassirQAS and Quran-BERT \citep{quranbert2021}. Wahid \citep{wahid2025} evaluates chatbot accuracy in reproducing Quranic interpretations, noting both the potential for democratizing access and the risk of hallucination. Sholeh and Yunusy \citep{sholeh2024} analyze the use of AI chatbots in religious education, finding that while students engage positively, educators remain cautious about delegating authority to machines. Nuraeni and Khairudin \citep{nuraeni2025} examine WhatsApp-based Islamic chatbots in Indonesia, highlighting their popularity for informal learning but also pointing to issues of oversimplification.  Shamsuddin \citep{shamsuddin2024} outlines the conditions under which chatbots might be considered halal, emphasizing transparency and the avoidance of authoritative rulings. 

Latifi \citep{latifi2024} considers the implications of AI for Shia practices of \textit{ijtihād}, suggesting that LLMs risk undermining the interpretive authority of the scholars. Abdelnour \citep{abdelnour2025} offers a broader theology of technology, arguing that the automation of religious knowledge challenges long-standing religious hierarchies. The risks of algorithmic bias in LLMs are well documented. Abid et al. \citep{abid2021} demonstrated persistent Islamophobic framings in GPT-3 outputs, where prompts about Muslims disproportionately generated associations with violence. Bunts work on Islamic cyberspace \citep{bunt2009} documents the early proliferation of fatwa websites and online Quran repositories. Campbell and Tsurias \textit{Digital Religion} \citep{campbell2021} provides a broader theoretical frame, showing how religious communities adapt and contest new media. 

\section{Algorithmic Authority, Ijtihad, and Epistemic Risk}
Islamic jurisprudence is traditionally grounded in a multi-layered epistemic structure that distinguishes between revealed sources (the Quran and Hadith), interpretive methodologies (\emph{usull al-fiqh}), and the human scholarly labor of reasoning (\emph{ijtihad}) \citep{quraishi2001principles}\citep{lowry2007early}. Within this tradition, disagreement (\emph{ikhtilaf}) is not treated as epistemic failure but as an expected and often productive outcome of principled interpretation wiithin reasonable bounds. The Islamic scholarly tradition recognizes legal plurality across schools of law (\emph{madhahib}), regions, and historical contexts. This reflects differences in evidentiary weighting, ethical reasoning, and methodological commitments rather than mere error.

LLM-based Islamic chatbots run the risk of being misaligned with classical juristic practice. Most such systems are optimized to generate a single fluent, coherent response, a design choice that tends to collapse legitimate juristic disagreement into an apparent consensus. Where human scholars situate rulings within explicit lineages of interpretation, methodological commitments, and scholarly debate, LLMs present outputs without an accountable interpretive stance or transparent chains of reasoning or transmission\emph{isnad}. A potential long term impact of proliferation of Islamic LLMs may be that one opinion may be implicitly elevated as normative while alternative authoritative positions remain invisible.

A further epistemic risk arises from the probabilistic nature of LLM outputs. Classical Islamic legal theory carefully distinguishes between definitive (\emph{qati}) and presumptive (\emph{zanni}) evidence \citep{ramadan2006ijtihad}. Scholarly conclusions are often accompanied by explicit caveats regarding uncertainty, context, or minority views. LLMs can easily conflate distinct forms of uncertainty: ambiguity arising from limited or noisy training data becomes indistinguishable from genuine juristic disagreement. From the user's perspective, a hallucinated response, a weak narration, and a contested but well-established legal position may appear equally authoritative.

Personalization further complicates these risks. By tailoring outputs to user preferences, linguistic style, or inferred beliefs, Islamic chatbots may inadvertently reinforce interpretive echo chambers. Systems optimized for engagement may preferentially surface views aligned with prior interactions rather than exposing users to the breadth of Islamic legal and theological thought. This personalization stands in tension with pedagogical traditions that emphasize disciplined study, sustained engagement with disagreement, and ethical restraint in issuing normative judgments.

Finally, the prospect of automated or semi-automated fatwa-like responses raises questions of moral responsibility. In Islamic ethics, issuing a legal ruling is a morally weighty act that entails accountability before both the community and God. When normative guidance is mediated through algorithmic systems, responsibility becomes diffused across developers, institutions, data curators, and models themselves. Without clear governance structures and mechanisms for correction, users may increasingly rely on chatbots for moral guidance while lacking avenues for contestation or appeal. Addressing these epistemic risks therefore requires not only technical safeguards but careful attention to how authority, uncertainty, and responsibility are represented in algorithmic religious systems.

\section{The Landscape of Islamic Chatbots}
Over the last few years, there has been a proliferation of Islamic chatbots and LLM-based assistants, ranging from state-backed initiatives to grassroots, open-source projects. These systems vary widely in their orientation, underlying technology, and degree of transparency. Some are explicitly positioned as tools for Quran or Hadith study and others as guidance platforms. These initiatives resemble earlier grassroots Islamic forums of the 1990s and 2000s, where lay Muslims experimented with new forms of digital authority beyond institutional control \citep{bunt2009}. In terms of systems developed by governments, among the most prominent examples is \textbf{HUMAIN Chat}, developed under Saudi Arabia's King Abdulaziz Center for World Culture (Ithra). It is powered by the ALLAM-34B model. While strictly not focused on religious knowledge, it is trained on Arabic and Islamic corpora and supposed to reflect Islamic values\citep{humain2024}. By contrast, many commercial or app-based services (e.g., Ask AiDeen, IslamicGPT.com, Shaykh.AI) offer little information about datasets or model architectures. Ansari Chat represents a notable outlier in its open-source release of both backend and frontend code.   Other chatbots occupy the middle ground between these poles. 

\textbf{Ask AiDeen}, integrated into the widely used Muslim Pro app, functions as a semi-official assistant, citing Quranic and Hadith sources while remaining opaque about its underlying model. \textbf{WisQu}, is a chatbot developed for the Shia community. The denominational differentiation in the physical world are also reflected in the digital domain, continuing the fragmentation of online fatwa portals  by sect and school of thought\citep{hirschkind2006, bunt2009}. Meanwhile, platforms such as \textbf{Salam.chat}, \textbf{SheikhGPT}, and \textbf{Shaykh.AI} explicitly market themselves to younger, digitally native Muslims, offering bite-sized religious content in conversational form. In addition to general-purpose chatbots, there are specialized systems focusing on textual corpora. \textbf{QuranGPT} and \textbf{MyQuran.online} restrict their scope to Quranic interpretation, while \textbf{HadithGPT} provides search and explanation of Prophetic traditions.  Similarly, \textbf{Tarteel}, initially a Quran memorization and recitation feedback app, now includes LLM-powered "Ask" features, extending the tradition of pedagogical technology into conversational interfaces. Another point of differentiation is scope. General-purpose assistants such as SheikhGPT or Shaykh.AI aim to provide fatwa-like responses across a wide range of topics. Specialized platforms such as QuranGPT and HadithGPT limit themselves to scriptural corpora, offering computational searchability.  A comparison summary of some of the more popular LLM powered chatbots is given in Table \ref{tab:islamic_chatbots}.

Institutional fatwa authorities have increasingly experimented with automation. A well-publicized early example came from Egypt in 2017, when Al-Azhar piloted staffed "fatwa kiosks" in Cairo provide quick, on-site guidance \cite{Reuters2017AlAzharKiosk,Guardian2017Kiosk}. In Dubai, the Islamic Affairs \& Charitable Activities Department (IACAD) launched ''Virtual Ifta'', an AI system that answered a fixed set of common questions via chat. IACAD and the UAE Fatwa Council have since expanded multi-channel ''chatbot fatwas'' alongside website, app, WhatsApp, and SMS services \cite{GulfNews2019VirtualIfta,UAEFatwaCouncilServices}. Saudi authorities introduced mobile ''fatwa robots'' inside the Grand Mosque complex, aiming to assist pilgrims at scale with multilingual guidance and even video calling to human scholars, a hybrid model that pairs automation with clerical oversight \cite{SPA2025Manara,ArabNews2024GuidanceRobot,InterestingEng2025Manara}. 

\begin{table}[t]
\caption{Islamic Chatbot Comparison}
\label{tab:islamic_chatbots}
\centering
\begin{tabularx}{\textwidth}{l X X}
\toprule
\textbf{Tool / Chatbot} & \textbf{Source Transparency} & \textbf{Key Features} \\
\midrule
\href{https://ansari.chat}{Ansari Chat} & RAG; Quran/Hadith & Quran/Hadith Q\&A; fiqh basics; multiple scholarly views; Arabic morphology; dua; stories; poetry \\
\href{https://www.humain.ai}{HUMAIN Chat} & ALLAM-34B & Dialect speech input, real-time search, bilingual, conversation sharing \\
\href{https://www.muslimpro.com}{Ask AiDeen} & Quran/Hadith cited & Integrated into Muslim Pro; fatwa-style guidance \\
\href{https://www.islamicity.org}{ChatILM (IslamiCity)} & Quran/ Hadith cited & Hadith search; civic and Quran-based Q\&A \\
\href{https://salam.chat}{Salam.chat} & Quran/Hadith references & Scholar-trained dataset; halal filters \\
\href{https://sheikhgpt.org}{SheikhGPT} & Minimal disclosure & Virtual scholar persona; rulings/explanations \\
\href{https://wisqu.ai}{WisQu} & Quran, Hadith, Shia fiqh & Jurisprudential focus; multiple sources \\
\href{https://fatwauae.gov.ae}{UAE Fatwa Council Chatbot} & 
State-backed; fixed authoritative corpus &  Automated responses to common fatwa questions\\
\href{https://quranly.app}{Quranly AI} & 
Quran-only; proprietary &  Verse exploration; scripture-constrained interaction \\
\href{https://myquran.online}{MyQuran.online} & Built on OpenAI GPT & Quran-only answers; privacy focus \\
\href{https://chatilm.islamicity.org/}{ChatILM (IslamiCity)} & 
Quran/Hadith cited; institutional content & 
Conversational Islamic Q\&A; Quran-based explanations; non-fatwa framing \\
\href{https://www.qurangpt.com}{QuranGPT} & GPT-3.5 Turbo disclosed & Quran-focused answers \\
\href{https://hadithgpt.com}{HadithGPT} & Trained on $\sim$40k hadith & Direct hadith search \\
\href{https://www.imamai.app}{ImamAI} & Some references to sources & ``QuranAI'' integrated into prayer app \\
\href{https://askquran.chat}{AskQuran.chat} & Claims citations & Quran/Hadith reference Q\&A \\
\href{https://maarifa.ai}{Maarifa AI} & None & Ramadan spiritual guidance \\
\href{https://islamicgpt.com}{IslamicGPT} & None & General Islamic guidance \\
\href{https://tarteel.ai}{Tarteel} & ASR engine disclosed & Recitation feedback; ``Ask'' beta \\

\bottomrule
\end{tabularx}
\end{table}

\section{Key Themes, Challenges \& Recommendations}
The deployment of LLMs in Islamic contexts raises a number of challenges. As with the introduction earlier technologies like the printing press, cassette sermons, and online fatwa portals, Islamic chatbots promise to democratize access to religious knowledge by lowering linguistic, financial, and institutional barriers. However, this democratization risks diluting the depth and rigor of scholarship. LLMs may provide instant answers, but they often reduce complex interpretive traditions into simplified or context-free responses. This mirrors earlier critiques of cassette sermons, which were accused of privileging emotional immediacy over scholarly precision \citep{hirschkind2006}. Another emerging issue is the decentering of religious authority. Historically, the rise of print and later broadcast media disrupted the monopoly of the \textit{ulama}, enabling charismatic preachers and the public to redefine Islamic discourse \citep{messick1993, armbrust1996}. Islamic chatbots reproduce this dynamic by offering fatwa-like guidance outside traditional institutions. Some, like HUMAIN Chat, seek to align AI outputs with official discourse, while others, like Ansari Chat, foreground open-source transparency. The question of ''who speaks for Islam'' becomes further complicated when the speaker is a probabilistic model trained on large datasets. Islamic chatbots also reproduce the sectarian pluralism of Muslim societies e.g., WisQu explicitly grounds itself in Shia thought, while most other chatbots adopt Sunni thought. While denominational specificity can enhance trust for particular communities, it also risks reinforcing polarization, particularly if LLMs are optimized to serve distinct sectarian markets rather than fostering cross-sectarian literacy.

One of the most pressing technical challenges for LLM-powered chatbots is scriptural grounding and source control. The Quran and authentic Hadith are central sources of authority. However, large language models,trained on broad, often uncurated corpora, can easily hallucinate Quranic verses, misquote Prophetic traditions, or conflate weak (daif) narrations with strong (sahih) ones. These errors risk not only spreading misinformation but violating the rigor with which Muslims approach revealed texts. Additionally, given the multiplicity of translations and commentaries, each with their own interpretive traditions, chatbots must implement version control and clearly indicate which scholarly authority or textual tradition underpins a given answer. 

Islamic chatbots inherit the broader problem of algorithmic bias in LLMs. Studies have shown that general-purpose models such as GPT-3 reproduce Islamophobic associations, disproportionately linking Muslims to violence or extremism \citep{abid2021}. Without careful dataset curation and evaluation, such biases may seep into Islamic chatbots, even when trained on religious corpora. This raises issues of epistemic justice: whose knowledge is encoded, whose is excluded. The opacity of large-scale models makes it difficult for users to discern whether outputs are grounded in canonical sources or shaped by latent biases in training data. Data provenances appears to be a paramount concern for Islamic LLMs. These are often grounded in retrieval-augmented generation (RAG) pipelines that ground responses in verifiable sources, especially Quran and Hadith. Outputs must be accompanied by citations, enabling users to trace responses to canonical texts.

Transparency is critical to building trust. We recommend developers to disclose model architectures, training data sources, and moderation protocols, in line with the broader calls for a transparent and accountable AI \citep{campbell2021}. Transparency also entails making clear disclaimers about the scope of the system i.e., distinguishing between educational guidance and a fatwa. Islam is a global religion characterized by linguistic diversity, sectarian pluralism, and cultural variety. Responsible Islamic LLMs should therefore be multilingual, and whenever possible inclusive of various jurisprudential traditions. LLMs should not replace scholars but serve as tools to augment their work. Human-in-the-loop mechanisms can include advisory boards of Islamic jurists, review processes for model updates, and explicit pathways for scholarly correction of erroneous outputs. This aligns with Islamic traditions of scholarly consensus (\textit{ijma}) and ongoing interpretation (\textit{ijtihad}), while also addressing contemporary AI concerns around accountability. Hybrid models that pair automation with clerical oversight may help reconcile accessibility with legitimacy. To summarize, the main challenges that LLMs need to address and some recommendations are as follows:
\begin{itemize}

\item \textbf{Scriptural Fidelity}: 
Islamic chatbots must ensure accurate quotation and attribution of Quranic verses and Hadith, including clear differentiation between authentic (\textit{sahih}), good (\textit{hasan}), and weak (\textit{da‘if}) narrations. Systems should avoid paraphrasing revealed texts in ways that obscure their original wording or context. Where interpretive commentary is provided, it should be explicitly distinguished from the primary text.

\item \textbf{Citation Transparency}: 
All generated responses should be traceable to identifiable canonical sources, such as specific Quranic verses, Hadith collections, or recognized scholarly works. Citations should be presented in a way that allows users to independently verify claims and consult the original texts. Lack of citation should be clearly indicated rather than masked by fluent generative output.

\item \textbf{Sectarian Drift}: 
Chatbots should avoid unintentionally blending jurisprudential traditions (e.g., Sunni and Shia fiqh, or distinct schools within Sunni Islam) without disclosure. When multiple authoritative opinions exist, systems should surface this plurality rather than defaulting to a single synthesized answer. Explicit labeling of jurisprudential context can help prevent misrepresentation of doctrinal positions.

\item \textbf{Bias Leakage}: 
Developers must actively audit Islamic chatbots for the reproduction of Islamophobic tropes, cultural stereotypes, or securitized framings inherited from general-purpose LLMs. Bias mitigation strategies should include curated religious corpora, targeted evaluation prompts, and ongoing monitoring of outputs. Given the moral and social stakes of religious guidance, even subtle bias warrants serious attention.

\item \textbf{Translation Fidelity}: 
Because Islamic knowledge circulates across multiple languages, chatbots should preserve semantic and theological consistency across translations. Key religious terms should not be flattened or mistranslated in ways that alter doctrinal meaning. Where multiple translations or interpretations exist, systems should acknowledge these differences rather than presenting a single definitive rendering.
\end{itemize}

\section{Conclusion \& Future Possibilities}
The use of religious chatbots powered by LLMs have proliferated in the last two years or so. Islamic chatbots are no exception to this phenomenon. The study of Islamic chatbots contributes not only to understanding Muslim societies but also to broader discussions about religion and technology. The use of Islamic chatbots powered by LLMs has implications for questions of authority since people may be tempted to ask religious questions to an LLM rather than ask a scholar or consult a book. LLMs thus enable highly personalized forms of religious interaction.  In the future, it may be possible that rather than fully displacing the \textit{ulama}, Islamic LLMs may evolve into tools that augment scholarly practice.  The proliferation of multiple denominational or movement-specific LLMs may foster algorithmic pluralism. Sunni, Shia, Sufi, or Salafi communities are already developing tailored chatbots, preserving doctrinal distinctions in digital form. While this may increase trust among users, it also risks deepening sectarian divides, as believers retreat into algorithmically mediated echo chambers. One could even speculate that we may even witness the emergence of AI ''muftis'' in the near future issuing rulings at scale, AI companions embedded into prayer apps and mosque infrastructures.

\bibliographystyle{plainnat}   
\bibliography{references}      

\end{document}